\documentclass[11pt]{article}

\usepackage{euscript}
\usepackage{amssymb}
\usepackage{amsfonts}
\usepackage{amsbsy}
\usepackage{amsmath}
\usepackage{epsfig}
\usepackage{amsthm}
\usepackage{amscd}
\usepackage{amstext}
\usepackage{verbatim}

\usepackage[pdftex]{hyperref}

\def\ben{\begin{equation}}
\def\een{\end{equation}}
\def\half{{\textstyle{1\over2}}}

 \let\b=\beta   
   \let\k=\kappa
 \let\m=\mu    
\let\s=\sigma

\let\w=\omega

\let\pa=\partial
\def\be{\begin{equation}}
\def\ee{\end{equation}}
\def\beq{\begin{equation}}
\def\eeq{\end{equation}}
\def\ba{\begin{array}}
\def\ea{\end{array}}

\def\dalemb#1#2{{\vbox{\hrule height .#2pt
       \hbox{\vrule width.#2pt height#1pt \kern#1pt
               \vrule width.#2pt}
       \hrule height.#2pt}}}

\newcommand{\bea}{\begin{eqnarray}}
\newcommand{\eea}{\end{eqnarray}}

\numberwithin{equation}{section}

\begin{document}

\begin{center}

{ \LARGE {\bf Holographically smeared Fermi surface:}} \\ {\Large {\bf Quantum oscillations and Luttinger count in electron stars}}

\vspace{1cm}

{\large Sean A. Hartnoll, Diego M. Hofman and Alireza Tavanfar

\vspace{0.7cm}

{\it Center for the Fundamental Laws of Nature, \\
Department of Physics, Harvard University,\\
Cambridge, MA 02138, USA \\} }

\vspace{1.6cm}

\end{center}

\begin{abstract}

We apply a small magnetic field to strongly interacting matter with a gravity dual description as an electron star.
These systems are both metallic and quantum critical at low energies. The resulting quantum oscillations are shown
to be of the Kosevich-Lifshitz form characteristic of Fermi liquid theory. It is seen that only fermions at a single radius
in the electron star contribute to the oscillations. We proceed to show that the Fermi surface area extracted from
the quantum oscillations does not obey the simplest statement of the Luttinger theorem, that is, it is not universally proportional
to the total charge density. It follows that our system is a non-Fermi liquid that nonetheless exhibits Kosevich-Lifshitz quantum oscillations.
We explain how the Luttinger count is recovered via a field theoretic description involving a continuum of `smeared' fermionic excitations.

\end{abstract}

\pagebreak
\setcounter{page}{1}

\section{Introduction}

Fermi liquid theory is a robust and weakly interacting low energy description of a finite density of interacting fermions \cite{Polchinski:1992ed, shankar}. In order to achieve non-Fermi liquid metallic behaviour, the low energy theory must contain additional gapless degrees of freedom beyond the particle-hole excitations of the Fermi surface. Gapless bosonic modes can arise either when a Fermi liquid is tuned to a quantum critical point \cite{hertz, millis, review} or in critical phases wherein kinematic constraints require the introduction of gauge symmetries and the associated protected gapless excitations \cite{anderson,mottreview}.

In non-Fermi liquids, the gapless bosonic modes generate relevant perturbations of Fermi liquid theory. The system therefore flows to a strongly interacting low energy regime. It has recently been appreciated that expansions in a large number $N$ of fermion species fail to halt the flow to strong coupling \cite{sungsik, Metlitski:2010pd, Metlitski:2010vm}. Cousins of the usual $(4-\epsilon)$-type expansions lead to controlled weakly interacting non-Fermi liquid fixed points \cite{nayak1,nayak2,mross}. In this letter  we aim to gain insight from an approach to metallic quantum criticality that is not built around weakly interacting notions of single particle Green's functions and self energies.

The holographic correspondence provides a classical dual description of certain strongly interacting large $N$ gauge theories. The most established examples are maximally supersymmetric Yang-Mills theories \cite{Maldacena:1997re}. More generally, one expects gravity duals for a `landscape' of field theories \cite{Denef:2009tp}. Known examples typically have $SU(N)$ gauge fields coupled to fermion and bosonic matter that can in addition be charged under global symmetries.
In these cases we can place the theory at a nonzero chemical potential $\mu$ for a global $U(1)$ symmetry and flow to energy scales $E \ll \mu$ \cite{Hartnoll:2009sz}. If the low energy theory contains a charged bosonic operator with low enough scaling dimension, then a superfluid phase transition will result \cite{Gubser:2008px, Hartnoll:2008vx, Denef:2009tp}. Alternatively, if the theory contains low lying charged fermionic operators, the `bulk' gravitational description is given by a fermion fluid that drives the theory to an IR fixed point characterised by a finite dynamical critical exponent $z$ \cite{Hartnoll:2009ns, Hartnoll:2010gu}.\footnote{
If neither sufficiently low lying bosons nor fermions are present in the operator spectrum, the bulk is an extremal black hole. Extremal black holes have an $AdS_2 \times R^2$ near horizon geometry and therefore correspond to $z=\infty$ fixed points. Phenomenologically attractive consequences of such fixed points have recently been emphasized \cite{MITagain, Sachdev:2010uj}. The zero temperature entropy density necessarily associated with $z=\infty$ scaling suggests that this feature is likely an artifact of the large $N$ limit. While the AdS$_2$ regime can be pushed to parametrically low energy scales by coupling the bulk Maxwell field to dilaton fields \cite{Taylor:2008tg, Goldstein:2009cv, Charmousis:2010zz}, eventually higher derivative terms in the action become important and are expected to stabilise the flow at a $z=\infty$ IR fixed point \cite{Sen:2005wa}.}

\begin{figure}[h]
\begin{center}
\includegraphics[height=220pt]{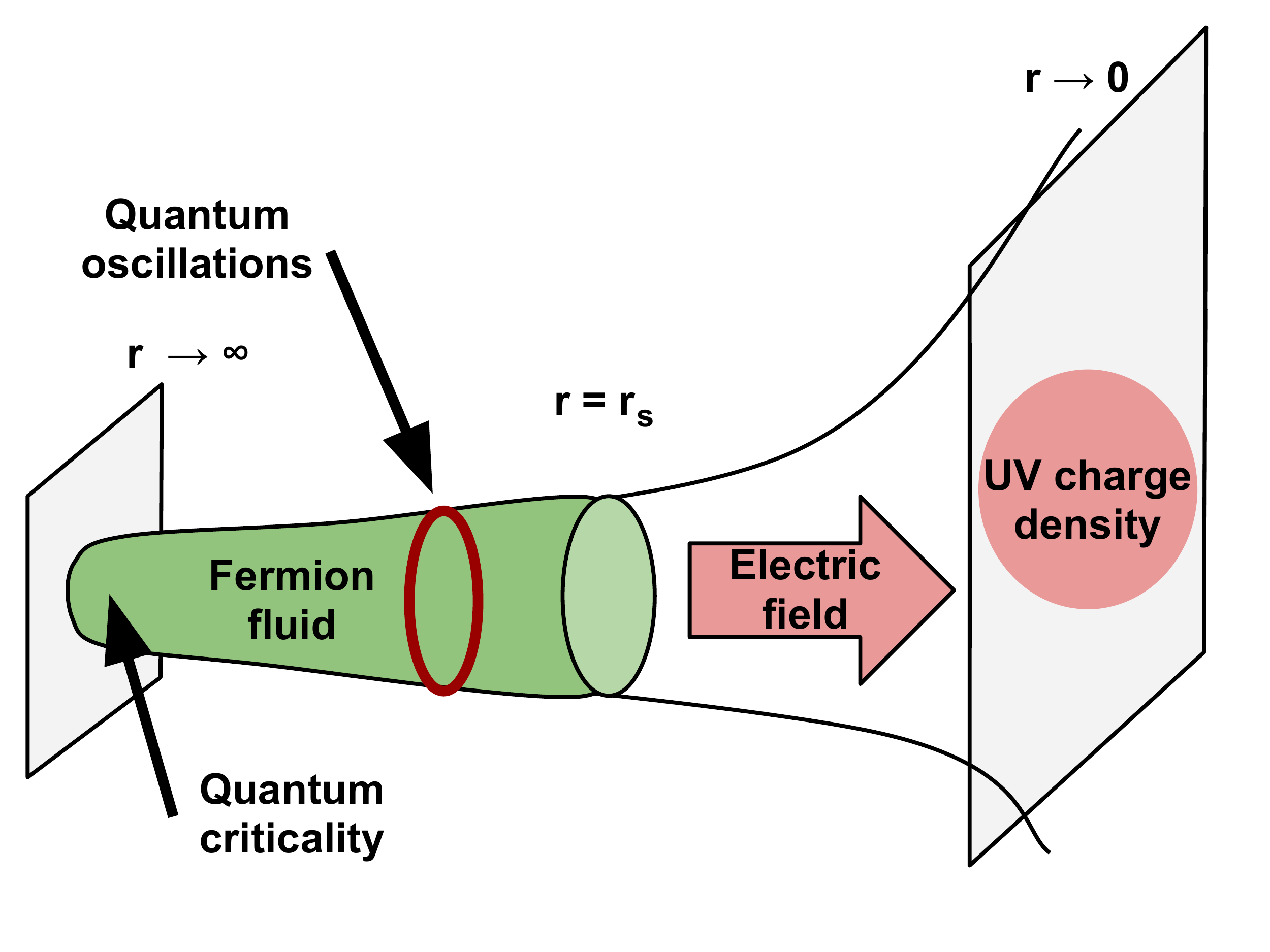}\caption{The electron star. The field theory charge density is given by the bulk electric field, which is sourced entirely by the fermion fluid. Quantum oscillations are due to fermions at a single radius in the star. The quantum critical region emerges due to screening of the electric field by the fermion fluid.\label{fig:picture}}
\end{center}
\end{figure}

A self gravitating fluid of charged fermions in asymptotically Anti-de Sitter space was termed an `electron star' in \cite{Hartnoll:2010gu} by analogy with the familiar neutron stars of astrophysics  \cite{openn, tolman}, which satisfy very similar equations. At zero temperature the entire charge is carried by the fermions. The absence of additional charge-carrying sectors of the theory, such as extremal black holes, suggests that electron stars are a promising arena to investigate the nature of strongly interacting Fermi surfaces. In this letter we will probe two key features of Fermi surfaces: their manifestation as quantum oscillations in a magnetic field and the Luttinger theorem relating the area of the Fermi surface to the total charge density \cite{oshikawa}. One of our main results will be that while the system exhibits the classic Kosevich-Lifshitz quantum oscillations characteristic of Fermi liquids, the corresponding Fermi surface area does not satisfy the Luttinger theorem. While a direct comparison is premature, it is noteworthy that this same apparent dichotomy between Fermi liquid oscillations and non-Fermi liquid physics has haunted much recent discussion in the cuprates following the seminal experiments of \cite{taillefer}. The geometry behind the mismatch in our holographic setup will become apparent: many of the bulk fermions do not participate in the oscillations and therefore disrupt the `Luttinger count'. In the final section we give a field theoretic description of this phenomenon and compare with `fractionalized Fermi liquids' \cite{ssv1,ssv2}. A sketch of an electron star is given in figure \ref{fig:picture}.

\section{Electron stars distilled}

The electron stars of interest to us will be zero temperature solutions of Einstein-Maxwell theory with a negative cosmological constant and coupled to a perfect fluid of unit charge relativistic fermions. Throughout this paper we will consider a 3+1 dimensional bulk for 2+1 dimensional field theories. The equations of motion are
\be\label{eq:einstein}
R_{ab} - \frac{1}{2} g_{ab} R - \frac{3}{L^2} g_{ab} = \k^2 \left(\frac{1}{e^2}
\left(F_{ac} F_b{}^c - \frac{1}{4} g_{ab} F_{cd} F^{cd} \right) + T_{ab} \right) \,,
\ee
and
\be\label{eq:maxwell}
\nabla_a F^{ba} = e^2 J^b \,.
\ee
As usual the perfect fluid energy momentum tensor and current are
\be\label{eq:ff}
T_{ab} = (\rho + p) u_a u_b + p g_{ab} \,, \qquad J_a = \s u_a \,.
\ee
These two quantities must be conserved and the four velocity $u$ normalised so that $u^2 = -1$. The pressure $p$, energy density $\rho$ and charge density $\sigma$ are fields on spacetime. They are given in terms of a local equation of state by
\be\label{eq:flat}
\rho = \int_{m}^{\mu_\text{loc.}} E\, g(E) dE \,, \qquad \sigma = \int_{m}^{\mu_\text{loc.}} g(E) dE \,, \qquad - p = \rho - \mu_\text{loc.} \sigma \,,
\ee
where $m$ is the fermion mass, the local density of states is
\be
g(E) = \b E \sqrt{E^2 - m^2} \,,
\ee
with $\b$ an order one number, e.g. $\b = \pi^{-2}$ for two component spinors, and the local chemical potential
\be
\mu_\text{loc.} = \frac{A_t}{\sqrt{g_{tt}}} \,.
\ee

The electron star has the following metric and Maxwell field
\be\label{eq:metric}
ds^2 = L^2 \left(- f dt^2 + g dr^2 + \frac{1}{r^2} \left( dx^2 + dy^2 \right) \right) \,, \qquad A = \frac{e L}{\k} h dt \,.
\ee
Here $f,g,h$ are functions of the radial coordinate $r$. The pressure and energy and charge densities are also functions of $r$. The velocity has nonzero component $u^t = 1/(L \sqrt{f})$. The couplings can be scaled out of the equations of motion by writing
\be\label{eq:rescale}
p = \frac{1}{L^2 \k^2} \hat p \,, \qquad \rho = \frac{1}{L^2 \k^2} \hat \rho \,,\qquad \sigma = \frac{1}{e L^2 \k} \hat \sigma \,, \qquad \hat \b = \frac{e^4 L^2}{\k^2} \b \,, \qquad \hat m^2 = \frac{\k^2}{e^2} m^2 \,.
\ee
The quantities $\hat m$ and $\hat \b$ are the two free dimensionless parameters of the theory.

The resulting equations of motion for $\{f,g,h,\hat p, \hat \rho, \hat \sigma \}$ are written down and solved numerically in \cite{Hartnoll:2010gu}. The key features of the solution are that

\begin{itemize}
\item Outside of the star radius, $r < r_s$, there is no fluid present and the spacetime is that of an AdS$_4$ planar Reissner-Nordstrom black hole. Thus the solution is asymptotically characterised by a chemical potential $\hat \mu$, charge density $\hat Q$ and energy density $\hat E$. The hats over these dual field theory quantities indicate that rescalings analogous to (\ref{eq:rescale}) have been performed. These appear in the geometry as
\be\label{eq:RN}
f = \frac{c^2}{r^2} - \hat M r + \frac{r^2 \hat Q^2}{2} \,, \qquad g = \frac{c^2}{r^4 f} \,, \qquad h = \hat \mu - r \hat Q \,.
\ee
The constant $c$ keeps track of the normalisation of the time coordinate. Zero temperature together with scale invariance of the UV theory imply that $\hat E = \frac{2}{3} \hat \mu \hat Q$.

\item Far inside the star, $r \gg r_s$, the solution acquires a scale-invariant form characterised by a dynamical critical exponent $z$
\be
f = \frac{1}{r^{2z}} \,, \qquad g = \frac{g_\infty}{r^2} \,, \qquad h = \frac{h_\infty}{r^z} \,.
\ee
The exponent $z$ can be determined numerically in terms of $\hat m$ and $\hat \beta$. At a given fixed $\hat m$ the exponent grows unboundedly as $\hat \beta \to 0$ and monotonically tends to a constant as $\hat \beta \to \infty$. For all $\{ \hat m, \hat \beta\}$
\be
z \geq \frac{1}{1 - \hat m^2} \,.
\ee
Furthermore in order for the electron star to exist the fermion must not be too massive. Specifically $0 \leq \hat m < 1$. This also ensures that $z \geq 1$.
\end{itemize}

For the emergent scaling exponent $z$ to be order one rather than parametrically large \cite{Hartnoll:2009ns} while simultaneously requiring the classical bulk description to be valid, one needs \cite{Hartnoll:2010gu}
\be\label{eq:condition}
e^2 \sim \frac{\k}{L} \ll 1 \,.
\ee
Recall that $L$ is the cosmological constant scale, $e$ the Maxwell coupling and $\k^2$ the Newton constant. The condition (\ref{eq:condition}) can be rephrased as the following dual field theoretic relation. Two `central charges' can be defined from the zero temperature current $J$ and energy momentum tensor $T$ correlations: $\langle T T \rangle \sim c$ and $\langle JJ \rangle \sim k$. As usual holography requires $c \gg 1$. The additional condition (\ref{eq:condition}) becomes the statement that $c \sim k^2$. The current $J$ is normalised here such that the fermion operator has charge one.

In between the UV and deep IR geometries is the bulk of the electron star. The star is an equilibrated self gravitating system of fermions that carry the entire electric charge. In the remainder of this paper we proceed to a dual field theory characterisation of the fermionic physics of the intermediate region, geometrically associated with the field theory energy scale $E \sim \mu$.

\section{Magnetic field probe of electron stars}

A density of fermions leads to robust characteristic features in the magnetic field dependence of thermodynamic quantities. Expanding in small momenta about a point $k_0$ on the Fermi surface of a Fermi liquid leads to the Green's function
\be\label{eq:GFL}
G^{-1}_\text{F.L.}(\w,k) = - i \w + v^\star k_\perp + \frac{k_\parallel^2}{2 m^\star} \,.
\ee
Here $\w$ is the Euclidean frequency measured relative to the chemical potential $\mu$. The constants $v^\star$ and $m^\star$ are the (locally defined on the Fermi surface) Fermi velocity and effective electron mass. The momenta $k_\perp$ and $k_\parallel$ are the components of the momentum in directions perpendicular and parallel to the Fermi surface at the point $k_0$. The dispersion relation (\ref{eq:GFL}) about a Fermi surface leads to de Haas-van Alphen quantum oscillations in the magnetic susceptibility. Restricting for simplicity to a circular Fermi surface, the leading oscillation in the susceptibility per unit area is given by the Kosevich-Lifshitz formula \cite{KL}
\be\label{eq:KL}
\chi_\text{osc.} = - \frac{\pa^2 \Omega}{\pa B^2} \propto \frac{\mu^2 m^\star}{B^2} \frac{2 \pi^2 m^\star T/B}{\sinh (2 \pi^2 m^\star T/B)} \cos \frac{A_F}{B} \,.
\ee
Here $T$ is the temperature, $B$ the magnetic field and $A_F = \pi k_F^2$ is the area of the Fermi surface. The Fermi momentum $k_F = m^\star v^\star$. Thus from a measurement of the susceptibility one can determine the two coefficients in the dispersion relation (\ref{eq:GFL}).

We will now search for an oscillatory magnetic susceptibility of our electron stars. Adding fully fledged magnetic fields and temperature dependence is involved and will not be considered here. Instead, we will work in a regime where in our field theory $T, \sqrt{B} \ll \mu$. This is typically the experimentally relevant regime. In this limit we can ignore the backreaction of temperature and magnetic field on the spacetime geometry as only the very near horizon region will be affected. Thus we keep the radial profiles of the functions $\{f,g,h,\hat p, \hat \rho, \hat \sigma \}$ satisfying the equations of the previous section and obtained numerically in \cite{Hartnoll:2010gu}. The crucial consequence of the magnetic field is to collapse the local fermion spectrum into Landau levels labelled by $\ell =0,1,2...$. While $B$ is taken small, $\ell B$ must be kept finite. The splitting of the spectrum leads to non-analyticities in the magnetic field dependence that are then smeared out by the temperature -- the competition between these two effects leads to the formula (\ref{eq:KL}) in a Fermi liquid.

Thus to leading order at low temperatures and magnetic fields we can use the expression for the free energy density from \cite{Hartnoll:2010gu}
\be
\hat \Omega = - \frac{1}{3} \hat \mu \hat Q = - \frac{c \hat \mu}{3} \int_{r_s}^\infty \frac{\sqrt{g(s)}}{s^2} \hat \sigma(s) ds \,.
\ee
This formula simply expresses the charge density of the dual field theory as an integral over the charge density in the electron star. In the presence of a small magnetic field and temperature, the local charge density $\sigma$ in (\ref{eq:flat}) must now be written as (we drop from the outset the contribution of antiparticles, these do not see the Fermi surface and hence do not contribute to quantum oscillations) 
\be\label{eq:sB}
\sigma = \frac{\beta \, B_\text{loc.}}{2} \int_{-\infty}^\infty dp \sum_\ell \frac{1}{1 + e^{(E(\ell,p)-\mu_\text{loc.})/T_\text{loc.}}} \,,
\ee
where $E(\ell,p) = \sqrt{p^2 + m^2 + 2 \ell B_\text{loc.}}$. We have dropped a factor of $\half$ that appears in the $\ell=0$ term. The local temperature and magnetic field are related to the rescaled temperature and magnetic field of the dual field theory via
\be\label{eq:BT}
B_\text{loc.}(r) = \frac{F_{xy}}{\sqrt{g_{xx} g_{yy}}} = \frac{e^2}{\k^2} r^2 \hat B \,, \qquad
T_\text{loc.}(r) = \frac{T}{\sqrt{g_{tt}}} = \frac{e}{\k} \frac{\hat T}{\sqrt{f}} \,.
\ee
The first of these relations used the fact that in the absence of magnetic sources the magnetic flux $F_{xy}$ is constant. In the next few equations we drop the `loc.' subscript on $\mu, T$ and $B$ for clarity of expression.

A finite temperature affects electron stars in two ways. It leads to Fermi-Dirac terms in the sum over states as in (\ref{eq:sB}) and also results in the appearance of a black hole in the interior of the geometry. The black hole is classical in the bulk and thereby captures the leading finite temperature physics whereas the effects of Hawking radiation, which is responsible for exciting the fermion fluid, are suppressed in the classical limit \cite{inprogress}. Changes in the local geometry, however, are insufficient to smooth out the quantum oscillations and therefore the expression (\ref{eq:sB}) contains the leading effects of temperature for our purposes.

While performing the sums in (\ref{eq:sB}) is a standard computation, we will outline the method used in \cite{Hartnoll:2009kk} which extracts the oscillating part efficiently and transparently. We first perform a Poisson resummation of the 
sum over Landau levels
\be
\sigma = \frac{\beta \, B}{2} \sum_k \, \int_0^{\infty} d\ell \, \int_{-\infty}^{+\infty} dp \,  \frac{e^{2 \pi i k \ell}}{1+ e^{(E(\ell,p) - \mu)/T}} \,.
\ee
We then proceed to change integration variables from $\ell$ to the energy $E$ and also to decompose the Fermi-Dirac distribution into a sum over Matsubara frequencies (the latter involves adding a $\mu$-independent term that will be cancelled by the antiparticle contribution). Thus
\be
\sigma = - \frac{\b\, T}{2} \sum_{k,n} \, \int_{-\infty}^{+\infty}  dp \, \int_{\sqrt{p^2+m^2}}^{\infty} dE \, E   \frac{e^{\pi i k (E^2-p^2-m^2)/B}}{E- \mu \left(1+ i \frac{T}{\mu} 2 \pi \left(n+\frac{1}{2}\right) \right)} \,.
\ee
The integral over $E$ is performed by deforming the integration contour in such a way that the argument of the exponential is always real and decreasing at large energies. The only obstruction to doing this is the pole coming from the Matsubara frequencies. This obstruction is only present if $p^2 + m^2  < \mu^2$ and furthermore $n+\frac{1}{2}>0$ for $k>0$ or $n+\frac{1}{2}<0$ for $k<0$. The rotated contour integral
is manifestly non oscillating in $B$. The only oscillating contribution comes from the poles and is given to leading order in $\frac{T}{\mu}$ by
\be
\sigma_\text{osc.} = 2 \pi \b\, T \mu \,  \text{Im} \sum_{k,n>0} {e^{\pi i k (\mu^2-m^2)/B}} {e^{-4 \pi^2  k\left(n-\frac{1}{2} \right) T \mu/B}}  \, \int_{-\sqrt{\mu^2-m^2}}^{\sqrt{\mu^2-m^2}}  dp \,  {e^{-\pi i k p^2/B}}  \,.
\label{eq:final}
\ee

\begin{figure}[h]
\begin{center}
\includegraphics[height=180pt]{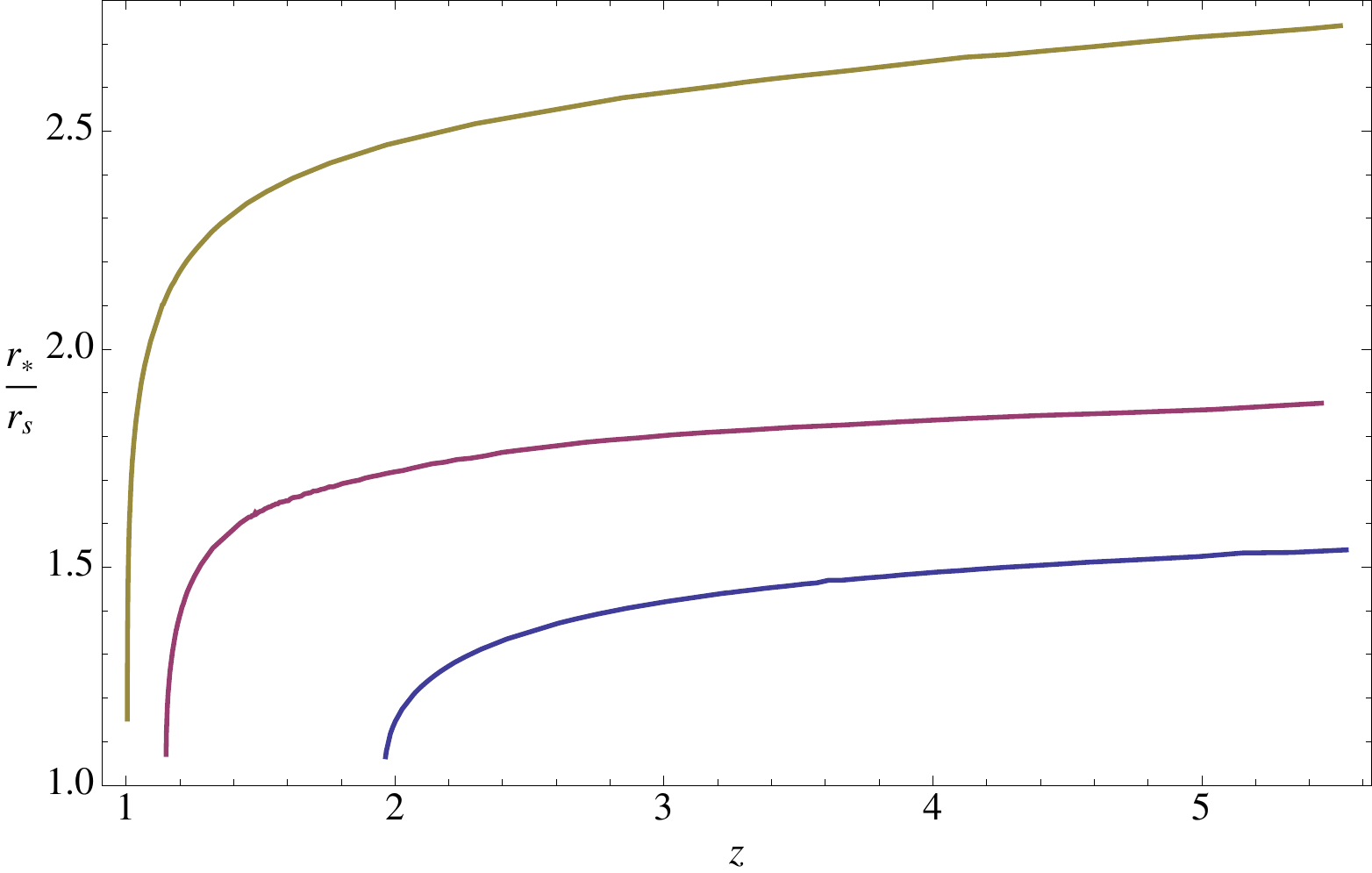}\caption{Radius contributing to quantum oscillations over the radius of the electron star. Larger $r_\star/r_s \geq 1$ is the interior of the star. From top to bottom, curves have $\hat m = 0.07, 0.36$ and $0.7$. We are characterising the solutions by their IR scaling exponent $z$ rather than $\hat \beta$. \label{fig:radii}}
\end{center}
\end{figure}

The expression (\ref{eq:final}) can now be plugged into the formula (\ref{eq:sB}) for the free energy of the boundary field theory. So long as the local Fermi surface area proportional to $\m^2 - m^2$ is not small, i.e. so long as we are not at the boundary of the electron star, then the fact that $\mu^2/B \ll 1$ allows us to perform a saddle point evaluation of the integral over radial position $s$ in (\ref{eq:sB}). Because $T \ll \mu$, the imaginary exponent in (\ref{eq:final}) in the more rapidly varying one. After taking the saddle point we restrict to the first harmonic $k=1$ as higher modes are suppressed and then perform the sum over the Matsubara frequencies $n$. Because $\m^2-m^2 \gg B$, the $p$ integral in (\ref{eq:final}) can be performed over the whole real line. The final result for the oscillatory free energy density is the Kosevich-Lifshitz form
\be\label{eq:KLstar}
\hat \Omega_\text{osc} = \frac{\hat \b\, r_\star \sqrt{g(r_\star)}}{3 \sqrt{2 \pi}} \frac{c \hat \mu \hat B^2}{ \sqrt{|\hat A_F''(r_\star)|}} \frac{2 \pi^2 \hat m^\star \hat T/\hat B}{\sinh (2 \pi^2 \hat m^\star \hat T/ \hat B)}  \cos \frac{\hat A_F}{\hat B} \,,
\ee
where we have expressed everything in terms of rescaled boundary field theory quantities. Comparing with (\ref{eq:KL}), the effective Fermi surface area is
\be\label{eq:AF}
\hat A_F = \max_{r} \frac{\pi (h^2/f - \hat m^2)}{r^2} \,,
\ee
with $r_\star$ being the radius at which the maximum is attained, and the effective fermion mass
\be\label{eq:mstar}
\hat m^\star = \left. \frac{h}{f r^2} \right|_{r=r_\star} \,.
\ee
The maximum required in (\ref{eq:AF}) exists and is unique. This is because the numerator, the local Fermi surface area, increases monotonically from zero at the boundary of the star to a constant in the interior, while the denominator is constant on the star boundary and diverges towards the interior $r \to \infty$. The physical reason the denominator appears is that the conserved magnetic flux is fixed at the boundary of spacetime and becomes concentrated towards the interior according to (\ref{eq:BT}). The radius $r_\star$ is shown in figure \ref{fig:radii} below for various values of the parameters $\{\hat m, z\}$, using the numerical methods of \cite{Hartnoll:2010gu}. The fact that low energy physics is captured at a bulk radius of order $\mu$ rather than $r^{-1} \ll \mu$ may indicate that the usual holographic IR/UV correspondence is complicated in the presence of a Fermi surface, analogously to the modified RG flow in field theory \cite{Polchinski:1992ed, shankar}.

The previous three equations are the first important result in this paper. We have shown that upon applying a small magnetic field at low temperatures, our strongly interacting field theory at finite density exhibits quantum oscillations with a Kosevich-Lifshitz form (\ref{eq:KLstar}). Furthermore, we found that the Fermi surface area and effective quasiparticle mass defined through these oscillations are determined by the properties of the electron star at a unique radius $r_\star$. The fact that only an extremal orbit in the Fermi surface contributes is of course familar \cite{onsager}.\footnote{This is seen clearly at zero temperature: A nonanalyticity in the free energy arises whenever a Landau level goes from being occupied at some radii to being above the local chemical potential at all radii. The nonanalyticity is therefore localised at the extremal radius (\ref{eq:AF}) at which the Landau level is last occupied.} The novelty in our setup is that the surface is extremal in real space rather than momentum space and that the spatial direction in question is not part of the field theory itself, but rather the emergent dimension characteristic of holographic duality. The general relativistic physics of the extra dimension manifests itself in e.g. the appearance of the metric function $f$ in (\ref{eq:mstar}): The effective mass and Fermi surface area are not related to each other as for free electrons with a chemical potential $\mu$, there is a gravitational redshift in the effective mass due to the fact that it is defined at a fixed radius in the bulk.

The dependence of $\hat A_F$ and $\hat m^\star$ on the parameters of the model $\{\hat m, z\}$ can be computed using the numerical techniques of \cite{Hartnoll:2010gu} and are shown in figure \ref{fig:ratios} below. It is important to plot dimensionless quantities and rescale by factors of $c$ in (\ref{eq:RN}) to ensure that theories are compared with the same time coordinate.
\begin{figure}[h]
\begin{center}
\includegraphics[height=130pt]{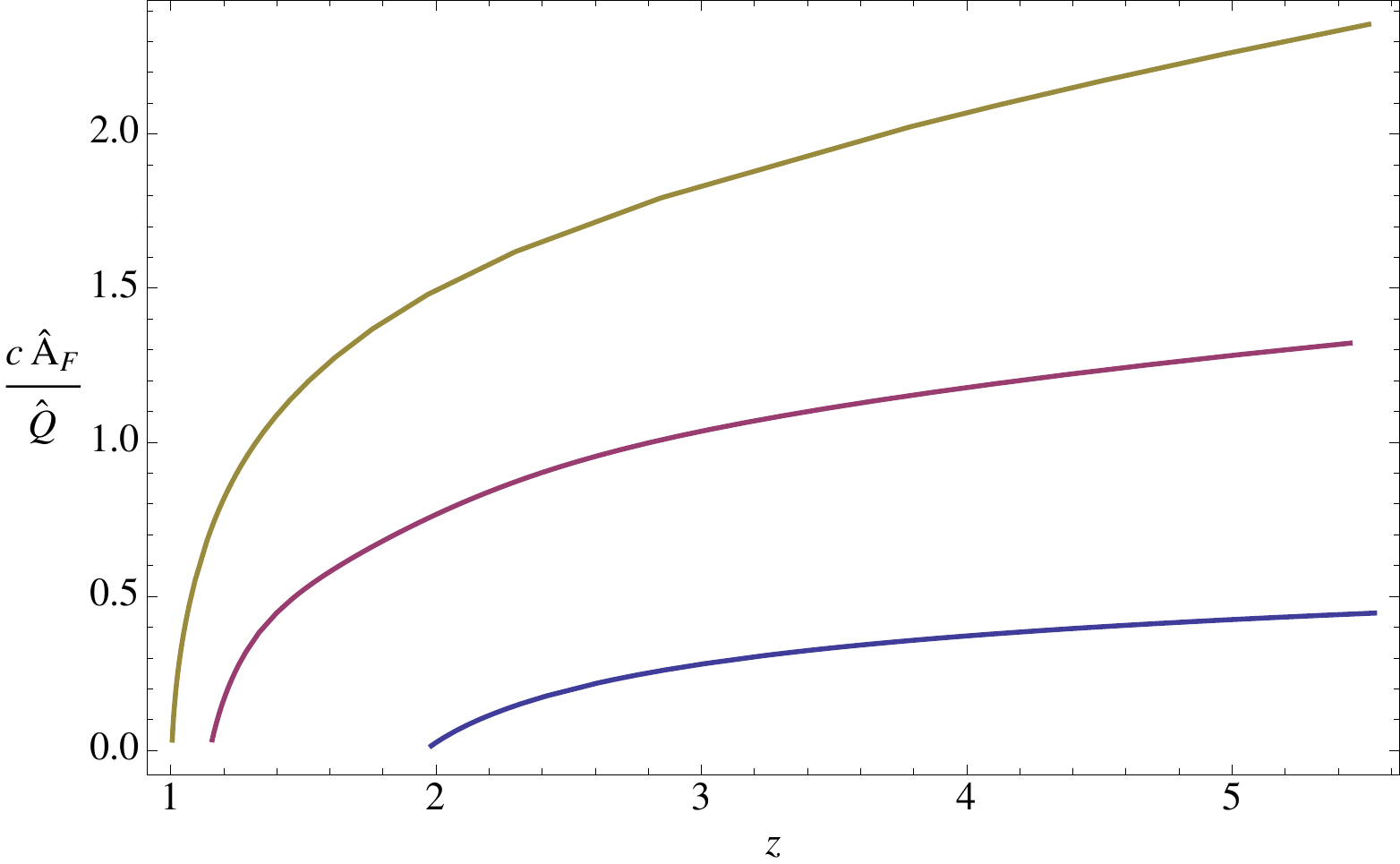}\hspace{0.3cm}\includegraphics[height=130pt]{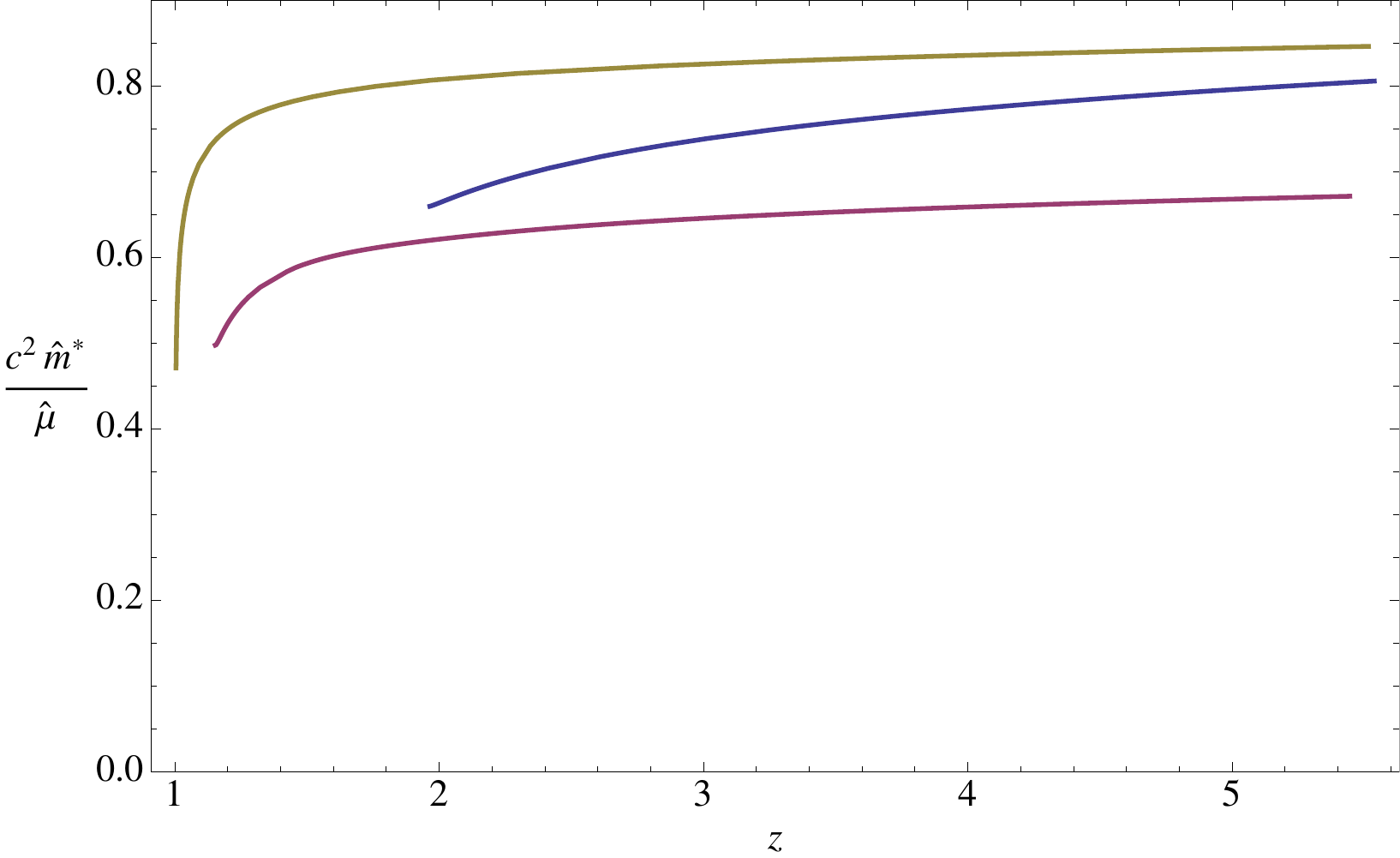}\caption{The `Luttinger ratio' of the Fermi surface area to charge density (left) and the effective quasiparticle mass over the chemical potential (right) as a function of IR scaling exponent $z$. From left to right, curves have $\hat m = 0.07, 0.36$ and $0.7$. \label{fig:ratios}}
\end{center}
\end{figure}
The left hand curve shows the Fermi surface area divided by the charge density. This is a dimensionless quantity without any factors of the speed of light. Restoring couplings $A_F/Q = \frac{e^3 L}{\k} \, \hat A_F/\hat Q \ll 1$. Although somewhat dependent on the convention for charge normalisation, this is consistent with the general holographic intuition that the field theory is in a `large $N$' limit, so that there many species of fermion and hence a large charge density with a fixed Fermi surface area. We call this quantity the `Luttinger ratio' as for a Fermi liquid the Luttinger theorem states that it is constant independently of the strength of interactions in the microscopic theory \cite{oshikawa}. The important conclusion of the left plot in figure \ref{fig:ratios} is that the Luttinger theorem does not hold for these systems, despite the fact that the quantum oscillations obey a Kosevich-Lifshitz form.\footnote{Previous holographic considerations of the Luttinger theorem appear in the context of probe fermions in an extremal black hole background \cite{Cubrovic:2009ye, Larsen:2010jt}. Without backreaction, the entire charge is carried by the black hole and the multiple probe Fermi surfaces include cases for which the quantum oscillations are not Kosevich-Lifshitz \cite{Denef:2009yy, Hartnoll:2009kk}. It is therefore not clear why it is appropriate to ask for a Luttinger theorem.} In the plot we have varied the UV dimension of the charged operator, via $\hat m$, and a ratio of the current and energy momentum central charges, via $\hat \beta$, see below (\ref{eq:condition}) above, keeping the charge of the fermionic operator constant. One might worry whether the normalisation of the charge operator should be changed simultaneously in a such a way that the Luttinger theorem holds. However, we will now explain the physics behind the violation of the Luttinger theorem in electron stars and their dual field theories.

\section{Field theory: Smeared Fermi surface and Landau bands}

From the bulk perspective it is clear what is going on. In the bulk we have a Fermi surface in a 3+1 dimensional spacetime. Because the spacetime is not homogeneous, the Fermi surface is not uniform in position space. As is well known \cite{onsager} and as we noted above, quantum oscillations are only sensitive to extremal cross sections of the Fermi surface. If a 3+1 dimensional Fermi surface is not simply a sphere in momentum space, then the area of the Fermi surface measured by quantum oscillations does not contain enough information to deduce the full volume of the Fermi surface. 
This is manifestly what is happening in our bulk setup: only fermions at an extremal radius are contributing to quantum oscillations. Hence the resulting Fermi surface area cannot determine the full volume of the bulk Fermi surface, which would need to be integrated over all radii. In our fluid limit the Luttinger theorem does hold locally at each radius. Therefore integrating the total Fermi volume over all radii would indeed reproduce the total charge density as required by the Luttinger theorem.

From the perspective of the dual strongly interacting field theory the interpretation is much more interesting. The dual field theory is 2+1 dimensional and spatially homogeneous. Therefore any Fermi surface will be a circle in momentum space and quantum oscillations should provide a unique definition of the area of this Fermi surface. The question becomes, in the spirit of \cite{oshikawa}, what low energy degrees of freedom are not captured by the quantum oscillations and are thereby responsible for violating the `Luttinger count'?

The bulk fermion field we have discussed so far corresponds to a `single trace' fermionic operator $\Psi$ in the dual field theory. In our `large $N$' limit, single trace operators have correlators that factorise despite the fact that they do not obey linear wave equations. Thus if we decompose the operator into energy and momentum modes
\be\label{eq:modes}
\Psi(x) = \int \frac{d\w d^2 k}{(2\pi)^3} \left(c_{\w,k} e^{- i \w t + i k \cdot x} + c_{\w,k}^\dagger e^{i \w t - i k \cdot x} \right) \,,
\ee
then we can expect a useful description in terms of the creation operators $c_{\w,k}^\dagger$.
Essentially the same decomposition is considered in \cite{deBoer:2009wk, Arsiwalla:2010bt}. Whereas those papers work on a spatial sphere and use the state-operator correspondence, we are using the existence of a Fock space at large $N$.
Unlike for free fields, however, we should not expect that only operators with a simple field theoretic `on shell' relation between $\w$ and $k$ contribute to physical quantities. Rather the on shell condition is that of the higher dimensional bulk theory
\be\label{eq:gauss}
g^{tt}(r) \w^2 + g^{xx}(r) k^2 + g^{rr}(r) k_r^2 = m^2 \,.
\ee
This shows that for a given field theory momentum $k$, fermions with a range of $\w$ will be important. We can subsume the two bulk labels\footnote{The presence of two labels is due to the coarse grained nature of the bulk ideal fluid description.} $k_r$ and $r$ into a single label, the effective mass $M$ of the field theoretic fermion. On shell fermions will then have
\be
\w_k(M) = \sqrt{c(M)^2 k^2 + M^2} \,.
\ee
In the limit in which we are working, bulk interactions are such that fermions modify the background geometry, but the fermions still satisfy the Gaussian dispersion (\ref{eq:gauss}).

We can now imagine integrating out all degrees of freedom except for the charge-carrying fermionic field $\Psi$.
The effective action that captures the fermionic ground state of the theory is then given by a continuum of modes labelled by
their mass $M$
\be\label{eq:Heff}
H_\text{eff.} = \int dM A(M) \int d^2k \left(\w_k(M) - \mu \right) c_k^\dagger(M) c_k(M) \,,
\ee
where in terms of the modes (\ref{eq:modes}) of the operator $\Psi$
\be
c_k(M) = c_{\w_k(M),k} \,.
\ee
The functions $A(M)$ and $c(M)$ in the action are determined in the process of integrating out the other fields. In the dual bulk description this amounts to solving the Einstein-Maxwell-fluid equations as was done in \cite{Hartnoll:2010gu}. The functions $A(M)$ and $c(M)$ can then be read off from the distribution of the fermion fluid in the bulk radial direction and the background spacetime geometry. The effects of the emergent IR scaling are contained in these functions. Because the single operator $\Psi$ does not lead to a single fermionic excitation but rather a continuum thereof, we might refer to the $c_k(M)$ excitations of (\ref{eq:Heff}) as `smeared' fermions.

Given the action (\ref{eq:Heff}), it is clear that the Luttinger count and quantum oscillations are reconciled in the same way as in the bulk description. The Fermi surface is labelled by both $M$ and the momentum. Only fermions that are at an extremal cross section in this total space contribute to quantum oscillations. Said differently, there is a continuum of Landau levels with an energy depending on $M$. These form `Landau bands'. Oscillations are due only to electrons at the extrema of the Landau bands. All of the fermions however contribute to the total charge density. Previous methods of violating the Luttinger theorem involve topological order, in which fractionalised excitations carry off part of the UV electronic charge density \cite{ssv1,ssv2}. Integrating out these excitations may lead to a structure similar to the effective smeared action (\ref{eq:Heff}). It would be very interesting to develop parallels. The holographic setup has the virtue that the smeared spectral weight can be described by (dually) free stable excitations.

The effective action (\ref{eq:Heff}) for the fermions does not capture all of the interesting low energy physics. In particular, the dynamical critical exponent $z$ in the IR can be thought of as being due to Landau damping of the massless gauge fields of the UV quantum field theory by the ground state density of fermions. We will elucidate this connection elsewhere.

\section*{Acknowledgements}

We thank Subir Sachdev for helpful conversations. Our research is partially supported by DOE grant DE-FG02-91ER40654, NSF grant PHY-0244821, the FQXi foundation and the Center for the Fundamental Laws of Nature.

\end{document}